\documentclass[a4paper,12pt]{article}
\usepackage{amssymb}
\usepackage{latexsym}
\usepackage[dvips]{graphicx}
\usepackage{ae} % or {zefonts}
\usepackage[T1]{fontenc}
\usepackage[ansinew]{inputenc}
\usepackage{amsmath}
\usepackage{graphicx}
\usepackage{color}
\newtheorem{thm}{Theorem}[section]

\newtheorem{prop}[thm]{Proposition}
\begin{document}
\begin{center}
{\LARGE{\bf On  Spherically Symmetric Non-Static Space-Times Admitting
 Homothetic Motions}}\\[1em]
\large{\bf{Ragab M. Gad}\footnote{Email Address: ragab2gad@hotmail.com}}\\
\normalsize {Mathematics Department, Faculty of Science,}\\
\normalsize  {Minia University, 61915 El-Minia,  EGYPT.}
\end{center}

\begin{abstract}
Spherically symmetric solutions admitting a homothetic Killing
vector field (HKVF) either orthogonal , $\eta_{\bot}$, or parallel,
$\eta_{||}$, to the 4-velocity vector field, $u^a$, are studied. New
self-similar solution of Einstein's field equation is found in the
case when HKVF is in a general form. Some physical properties of the
obtained solution are examined.
\end{abstract}

%%% ----------------------------------------------------------------------
%\maketitle
%%% ----------------------------------------------------------------------
\setcounter{equation}{0}
\section{Introduction}
Recently, symmetries in general relativity have attracted much
attention, not only because of their classical physical
significance, but also because they simplify Einstein field
equations. Many survey articles are given to discuss the concept of
these symmetries from the mathematical and physical viewpoints (see
for example \cite{H96})
\par
One of the most important symmetries is the self-similarity.
Self-similar solutions of the Einstein field equations are of great
interest in physics because they are often found to play an
important role in describing the asymptotic properties of more
general solutions \cite{HW86}. These solutions have relevance in
astrophysics and critical phenomena in gravitational collapse (see
for example \cite{CC99} - \cite{CC05} and references therein).
\par

 In a recent paper \cite{GH03}, Gad
and Hassan studied a non-static spherically symmetric solutions.
They assumed that these space-times admit a homothetic vector
field orthogonal to the 4-velocity vector, $u^a$, and obtained an exact
solution. This solution has  non-vanishing expansion, acceleration
and shear. They derived another solution by assuming, in additional
to space-like homothetic motion, the matter in this fluid is
represented by perfect fluid. This solution has zero expansion.
\par
Many exact solutions has been derived by imposing the condition of
the existence of a conformal Killing vector orthogonal to the
4-velocity (see for example  \cite{K94}, \cite{G02}). \\
In the present
paper we study the cases when space-time admitting HKVF either
orthogonal or parallel to 4-velocity vector. Several authors have
studied the solutions admitting the first symmetry. Most of them
have restricted their intention to the solutions discovered by
Gutman-Bosal'ke \cite{6}, which are given in another form by Wesson
\cite{10}. These solutions are denoted by (GBW). Collins and Land
\cite{4} have studied these solutions as well as a stiff equation of
state. Sussman \cite{9} investigated the properties of them, and
obtained interesting results.\\
The second aim of this paper is to obtain an exact self-similar
solution and explore some of its physical properties.
\par
The paper has been organized as follows: In the next section, we
shall comment on the singularities inherent the solutions obtained
in \cite{GH03} and we examine when these singularities could be
possible naked. We find the form of HKVF when it is either
orthogonal or parallel to $u^a$. We shall derive an exact new
self-similar solution. In section 3, we shall discuss the physical
properties of the obtained solution. Finally, in section 4, we shall
conclude the results.

\setcounter{equation}{0}
\section{\bf{Homothetic Motion}}
A global vector field $\eta$ on a space-time $M$ is called
homothetic if either one of the following conditions holds on a
local chart:
\begin{equation}\label{con}
\pounds_{\eta} g_{ab} = \eta_{a;b} + \eta_{b;a} = 2\Phi g_{ab},
\qquad H_{a;b} = \Phi g_{ab} + F_{ab},
\end{equation}
where $\Phi$ is a constant on $M$, $\pounds$ stands for the lie
derivative operator, a semi-colon denotes a covariant derivative
with respect to the metric connection, and $F_{ab} = - F_{ba}$ is
the so-called homothetic bivector. If $\Phi \neq 0$, $\eta$ is
called proper homothetic and if $\Phi = 0$, $\eta$ is called Killing
vector field on $M$.
\par
For a geometrical interpretation of (\ref{con}) we refer the reader
to \cite{5}, \cite{6}, and for a physical properties we refer for
example to \cite{CC99}.
\par
For the study of non-static spherically symmetric motion, we used the model
given by \cite{11}
\begin{equation}\label{met}
ds^2 = \alpha d\nu^2 + 2\beta d\nu dr - r^2 (d\theta^2 +
\sin^2\theta d\phi^2),
\end{equation}
where $\alpha$ and $\beta$ are positive function of $\nu$ and $r$.
\par
Gad and Hassan \cite{GH03} assumed an additional symmetry to the
spherically symmetric, space-like homothetic motion, and they
obtained
\begin{equation}\label{AB}
\alpha = r^2 h(\nu) \qquad \beta = rf(\nu),
\end{equation}
where $h(\nu)$ and $f(\nu)$ are arbitrary positive functions. In
additional to the above symmetry, they assumed that the matter is
represented by a perfect fluid and found the relation between $f(\nu)$
and $h(\nu)$ as follows
\begin{equation}\label{hf}
h(\nu) = \frac{1}{2}f^{2}(\nu).
\end{equation}
This solution is scalar-polynomial singular along $r=0$ \cite{GH03}.\\
In the following we examine when this singularity could be possible
naked.
 To do this, we consider the transverse radial null geodesics.
The equations governing these geodesics are
\begin{equation} \label{4.1.1}
f(\nu) \ddot{\nu} + (f^{\prime}(\nu) - h(\nu))\dot{\nu} = 0,
\end{equation}
\begin{equation} \label{4.1.2}
r^2 h(\nu)\dot{\nu}^{2} + 2rf(\nu)\dot{\nu}\dot{r} = 0.
\end{equation}
It is clear from equation (\ref{4.1.2}) that the ingoing null
geodesics are the line $\nu$ = constant. The outgoing geodesics obey
the equation
\begin{equation} \label{4.1.3}
\frac{dr}{d\nu} = - \frac{rh(\nu)}{2f(\nu)}.
\end{equation}
By integrating this equation, we get
\begin{equation} \label{4.1.4}
r = c_{1} \exp \big( - \int{\frac{h(\nu)}{2f(\nu)}}d\nu\big), \qquad
c_{1} \neq 0,
\end{equation}
we can see the structure of the space-time by examining equation
(\ref{4.1.4}). For example, if there exists a solution of equation
(\ref{4.1.4}) which starts from the singularity and ends at the
future null infinity, the singularity is globally naked.
Unfortunately, we cannot solve equation (\ref{4.1.4}) unless the
special choices of the functions $f(\nu)$ and $h(\nu)$ are given.
Now, we have two cases are depending on the value of integrand
$\int{\frac{h(\nu)}{f(\nu)}} d\nu$, inside equation (\ref{4.1.4}).
\begin{enumerate}
\item If the integrand  has negative values, then the geodesics
will never meet the singular point $r = 0$.
 \item If the integrand  has positive
values, then the geodesics are meeting the singular point $r = 0$
\end{enumerate}

\begin{prop}
All non-static spherically symmetric solutions described by metric
(\ref{met}) admit a homothetic vector field orthogonal to the
4-velocity vector in the form $\eta_{\bot} = \Phi r\partial_{r}$.
\end{prop}
{\bf{proof:}}
\par
 Consider the homothetic Killing equation (\ref{con})
and $\eta$ is a homothetic Killing vector field having the general form
\begin{equation}\label{HVF}
\eta = A(\nu, r) \partial_{\nu} + \Gamma (\nu, r)\partial_{r}.
\end{equation}
For the metric (\ref{met}), we have
$$
u^a = \frac{1}{\sqrt{\alpha(\nu, r)}}.
$$
If $\eta_{\bot}$ is everywhere orthogonal to $u^a$, then
$$
\eta^a_{\bot} = \Gamma(\nu, r)\delta^a_r
$$
By straightforward calculations, using (\ref{AB}) and the
Christoffel symbols of second kind (see Appendix), we get that this
vector satisfies the condition (\ref{con}) if $\Gamma(\nu, r) = \Phi
r$. \\

 According to the above proposition and using (\ref{hf}), the
following result has been established
\begin{prop}
All perfect fluid solutions described by the metric (\ref{met})
admit a homothetic vector field orthogonal to the 4-velocity vector
in the form $\eta_{\bot} = \Phi r\partial_{r}$.
\end{prop}
\par
Now, we study the case when the homothetic vector field is parallel
to the four-velocity vector field.
\begin{prop}
All non-static spherically symmetric solutions described by metric
(\ref{met}) admit a homothetic vector field parallel to the
4-velocity vector in the form $\eta_{||} = \Phi \nu\partial_{\nu}$.
\end{prop}
{\bf{proof:}}
\par
 Consider the general form of HVF (\ref{HVF}) and
using the relation,since HVF is parallel to $u^a$,
$$
\eta^{a}_{||} = const.u^a,
$$
then
$$
\eta^{a}_{||} = A(\nu,r)\delta^{a}_{\nu}
$$
By straightforward calculations, using (\ref{AB}) and the
Christoffel symbols of second kind (see Appendix), we get that this
vector satisfies the condition (\ref{con}) if $A(\nu, r) = \Phi
\nu$. \\

 By the same manner, see proposition (2.2), we can prove
that if the fluid is a perfect fluid, then it admits HVF parallel to
$u^a$ in the form $\eta_{||} = \Phi \nu\partial_{\nu}$.
\par
According to the above propositions, the HVF $\eta$ takes the
following form
\begin{equation}\label{HV}
\eta = \phi r\partial_{r} + \phi \nu \partial_{\nu}
\end{equation}
This vector satisfies the conditions (\ref{con}).
\par
 Now we assume that the line element (\ref{met}) admits HVF (\ref{HV}), then the
(non-trivial) equations arising from (\ref{con}), are
$$
\nu\beta_{\nu} + r\beta_{r} = 0,
$$
$$
r\alpha_{r} + \nu\alpha_{\nu} = 0.
$$
 Using equations  (\ref{AB}) and (\ref{hf}), the solutions of the above equations are
$$
\alpha =\frac{1}{2} (\frac{r}{\nu})^2,
$$
$$
\beta = (\frac{r}{\nu}).
$$
According to the above results, the line element (\ref{met}) can be
written in the following form
\begin{equation}\label{GAD}
ds^2 = \frac{1}{2}(\frac{r}{\nu})^2 d\nu^2 + 2(\frac{r}{\nu}) d\nu
dr - r^2 (d\theta^2 + \sin^2\theta d\phi^2)
\end{equation}
In the following section, we shall discuss  some of the physical
properties of the obtained solution given by (\ref{GAD}).
\section{Physical Properties}
\subsection{Kinematic of the Velocity Field}
For a given space-time the kinematics properties (acceleration,
expansion scalar, rotation, shear and scalar shear) are respectively
defined
as below \cite{6}:\\
The acceleration is defined by
$$
\dot{u}_a = u_{a;b}u^b.
$$
The expansion scalar, which determines the volume behavior of the
fluid, is defined by
$$
\Theta = u^a_{;a}.
$$
The rotation is given by
$$
\omega_{ab} = u_{[a;b]} + \dot{u}_{[a}u_{b]}.
$$
The shear tensor, which provides the distortion arising in a fluid
flow leaving the volume invariant, can be found by
$$
\sigma_{ab} = u_{(a;b)}+\dot{u}_{(a}u_{b)}- \frac{1}{3}\Theta
h_{ab},
$$
where $h_{ab}= g_{ab} + u_{a}u_{b}$.\\
The shear invariant is given by
$$
\sigma^2 =\frac{1}{2} \sigma_{ab}\sigma^{ab}.
$$
For the solution given by (\ref{GAD})\\
  The acceleration is
$$
\dot{u}_{a} = - \frac{1}{2r}\delta^{1}_{a}
$$
For the expansion scalar, we find
$$
\Theta = 0.
$$
The only non-vanishing components of rotation is given by
$$
\omega_{41} = \frac{1}{\sqrt{2}\nu}.
$$
  The only non-zero component of the shear tensor is
$$
\sigma_{11} = - \frac{2\sqrt{2}}{r},
$$

and the shear scalar,  is given by
$$
\sigma^{2} = \frac{1}{r^2}
$$
\subsection{Pressure and Density}
In addition to self-similarity, we assume that the matter is
represented by a perfect fluid, that is, the Einstein field
equations, $G_{ab}=-\kappa T_{ab}$, are satisfied with the energy
momentum tensor
$$
T_{ab} =(\rho +p)u_au_b -pg_{ab}.
$$
For the line element (\ref{GAD}), the Einstein field equations
reduce to the following equations
$$
\frac{1}{r^2}=\kappa (\rho +p),
$$
$$
\frac{1}{2r^2}=\kappa p.
$$
From the above equations, we obtain the expression for the pressure
and density in the form
$$
p=\rho = \frac{1}{2\kappa r^2}.
$$
\subsection{Tidal Forces}
The components of the Riemann curvature tensor $R^{a}_{bcd}$, which
describe tidal forces (relative acceleration) between two particles
in free fall, are the components
\, $R^{i}_{0j0}$, ($i, j = 1, 2, 3$), \cite{7}.\\
For the line element (\ref{GAD}), we obtain
$$
R^{1}_{010} = 0,
$$
and the only non-vanishing relevant components are
$$
R^{2}_{020} = R^{3}_{030} = \frac{1}{4\nu^2}.
$$
Then the equations of geodesic deviation (Jacobi equations), which
connected  the behavior of nearby particles and curvature,
are reduce to the following equations
\begin{equation}\label{4.1}
\frac{D^{2}\zeta^{r}}{d\tau^2} = 0,
\end{equation}
\begin{equation}\label{4.2}
\frac{D^{2}\zeta^{\theta}}{d\tau^2} =- \frac{1}{2r^{2}}
\zeta^{\theta},
\end{equation}
\begin{equation}\label{4.3}
\frac{D^{2}\zeta^{\phi}}{d\tau^2} =- \frac{1}{2r^{2}}\zeta^{\phi},
\end{equation}
where $\zeta^r , \ \zeta^{\theta} , \ \zeta^{\phi}$ are the components of Jacobi vector field.\\
Hence, equation (\ref{4.1}) indicates tidal forces in radial
direction will not stretch an observer falling in this fluid. The
equations (\ref{4.2})and (\ref{4.3}) are indicate a pressure or
compression  in the transverse directions, that is,
 the tidal forces   will not squeeze the observer in the transverse
directions.

\section{Conclusion}
In the theory of general relativity, there  are  different types of
self-similarity. To distinguish between them we refer the reader to
the topical review by Carr and Coley \cite{CC99}. In this paper we
have restricted our intention to the first type of self-similarity,
which characterized by the existence of a homothetic Killing vector
field. We have obtained the form of homothetic Killing vector field
when it is either orthogonal or parallel to the 4-velocity vector
field. In the case when HKVF takes a general form, we have derived
self-similar solution. This solution has zero expansion,
non-vanishing acceleration and  non-vanishing shear and satisfies
the equation of state $\rho = p$. Furthermore,  we have shown that
the tidal forces in radial direction will not stretch an observer
falling in this fluid and they not squeeze him in transverse
directions.

\section*{\bf{Appendix}}
We use $(x^0, x^1, x^2, x^3) = (\nu, r, \theta, \phi)$ so that the
non-vanishing Christoffel symbols of the second kind of the line
element (\ref{met}) are
\begin{align*}
\Gamma^{1}_{11} & =  \frac{\beta_{r}}{\beta}, & \Gamma^{2}_{12} & =  \frac{1}{r},  \\
\Gamma^{1}_{22} & = - \frac{r\alpha}{\beta^2}, & \Gamma^{3}_{13} & =  \frac{1}{r},  \\
\Gamma^{1}_{01} & =  \frac{\alpha_{r}}{2\beta}, & \Gamma^{2}_{33} & =  - \sin\theta\cos\theta,  \\
\Gamma^{1}_{00} & =  -\frac{\alpha(\beta_{\nu}- \frac{1}{2}
\alpha_{r})}{\beta^2} + \frac{\alpha_{\nu}}{2\beta},
& \Gamma^{3}_{23} & =  \cot\theta,  \\
\Gamma^{0}_{00} &=\frac{\beta_{\nu}-\frac{1}{2}\alpha_{r}}{\beta}.
\end{align*}

\end{document}